\documentclass[reqno,11pt]{amsart}
\usepackage{graphicx}
\usepackage{amssymb}
\usepackage{amsmath}
\usepackage{amsfonts}
\usepackage{amsthm}
\usepackage{color}
\usepackage{hyperref}
\usepackage[mathscr]{eucal}
\usepackage{enumerate}
\usepackage{times}
\usepackage{authblk}
\usepackage[foot]{amsaddr}

\numberwithin{equation}{section}
\allowdisplaybreaks[1]

\newtheorem{Def}{Definition}[section]
\newtheorem{Thm}[Def]{Theorem}

\newtheorem{Conj}[Def]{Conjecture}
\newcommand{\beq}{\begin{equation}}
\newcommand{\eeq}{\end{equation}}
\newcommand{\Proof}{\begin{proof}}
\newcommand{\QED}{\end{proof} \noindent}

\newcommand{\mm}{\hspace{-.08cm}\cdot \hspace{-.08cm}}

\newcommand{\M}{\mathcal{M}}

\newcommand{\R}{\mathbb{R}}

\newcommand{\Gammati}{\tilde{\Gamma}}

\title[Strong Cosmic Censorship with Bounded Curvature]{Strong Cosmic Censorship with Bounded Curvature}

\author[M.\ Reintjes]{Moritz Reintjes \\ \\ April 10, 2023}
\address{Department of Mathematics\\ City University of Hong Kong \\ Hong Kong SAR}
\email{moritzreintjes@gmail.com}

\begin{document}

\maketitle

\begin{abstract}
In this paper we propose a weaker version of Penrose's much heeded Strong Cosmic Censorship (SCC) conjecture, asserting inextentability of maximal Cauchy developments by manifolds with Lipschitz continuous Lorentzian metrics and Riemann curvature bounded in $L^p$. Lipschitz continuity is the threshold regularity for causal structures, and curvature bounds rule out infinite tidal accelerations, arguing for physical significance of this weaker SCC conjecture. The main result of this paper, under the assumption that no extensions exist with higher connection regularity $W^{1,p}_\text{loc}$, proves in the affirmative this \emph{SCC conjecture with bounded curvature} for $p$ sufficiently large, ($p>4$ with uniform bounds, $p>2$ without uniform bounds).
\end{abstract}

\setcounter{tocdepth}{1}

\section{Introduction}

In this paper we propose and address a weaker form of Penrose's {\it Strong Cosmic Censorship}  (SCC) conjecture, subject to bounded curvature. For this we consider solutions of the Einstein equations of General Relativity
\beq \label{Einstein_eqn}
R_{\mu\nu} - \frac12 R g_{\mu\nu} + \Lambda g_{\mu\nu} = 8\pi T_{\mu\nu}
\eeq
where $g_{\mu\nu}$ is a Lorentzian metric on spacetime, $R_{\mu\nu}$ its Ricci tensor, $R$ its scalar curvature, $T_{\mu\nu}$ the energy momentum tensor of matter fields and $\Lambda \in \R$ the cosmological constant \cite{Einstein}. We begin by stating Penrose's original Strong Cosmic Censorship conjecture \cite{Penrose}, which has been subject of many investigations over the last 25 years, (see  \cite{Christodoulou1, Dafermos, DafermosRodnianski, Ringstroem, Natario, Natario2, LukSibierski, Franzen1, Franzen2, Kehle, LukOh, DafermosLuk} and references therein).

\begin{Conj} \label{SSC_1} 
{\rm (Penrose's SSC Conjecture)} \\
For generic compact, asymptotically flat or asymptotically anti-de Sitter initial data, the maximal Cauchy development of \eqref{Einstein_eqn} is inextendable as a manifold with Lorentzian metric in $C^0$.
\end{Conj}

Identifying a suitable notion of ``genericity'' of data (beyond the absence of symmetries) is part of the problem, and recent work of Kehle shows validity of the SSC conjecture may depend on the notion of genericity \cite{Kehle}. For the purposes of this paper, we can leave the notion of genericity of data open, and we do not need to specify the matter fields in \eqref{Einstein_eqn}.  We refer from now on to generic compact, asymptotically flat or asymptotically anti-de Sitter data simply as {\it generic data}. 

Dafermos and Luk recently disproved Conjecture \ref{SSC_1} in the case $\Lambda \geq 0$, under the assumption that the exterior Kerr spacetime is stable \cite{DafermosLuk}. The following weaker form of Penrose's SSC conjecture, introduced by Christodoulou \cite{Christodoulou}, remains open.

\begin{Conj} \label{SSC_2} 
{\rm (Christodoulou's SSC Conjecture)} \\
For generic data, the maximal Cauchy development of \eqref{Einstein_eqn} is inextendible as a manifold with $C^0$ Lorentzian metric and metric connection $\Gamma$ locally in $L^2$.
\end{Conj}

The $W^{1,2}$ metric regularity of Conjecture \ref{SSC_2} is the minimal regularity for which weak solutions of the Einstein equations can be introduced, and is therefore a well motivated lower threshold regularity for physical significance of extensions. In this paper we propose the following weaker form of Conjectures \ref{SSC_1} and \ref{SSC_2}, addressing Lipschitz continuous ($C^{0,1}$) Lorentzian metrics subject to bounded curvature. 

\begin{Conj} \label{SSC_3}
{\rm (SSC Conjecture with $L^p$ Curvature Bound)} \\
For generic data, the maximal Cauchy development of \eqref{Einstein_eqn} is inextendible as a manifold with $C^{0,1}$ Lorentzian metric and Riemann curvature locally in $L^p$, $p\geq 2$.
\end{Conj}

Conjecture \ref{SSC_3} may be viewed as an intermediate step for establishing Conjecture \ref{SSC_1} or \ref{SSC_2}. However, Conjecture \ref{SSC_3} also entertains the possibility that suitable curvature bounds could serve as a criterion for physical relevance of extensions of maximal Cauchy developments, keeping in mind that boundedness of the Riemann curvature in $L^\infty$ rules out infinite tidal accelerations by the Jacobi equations of geodesic deviation. Moreover, Lipschitz continuity of Lorentzian metrics has been established by Chruciel and Grant as the minimal regularity for causal structures to be well-defined \cite[Thm 1.20]{ChrucielGrant}, which places Lipschitz continuity of Lorentzian metrics as a threshold regularity for physical significance.   The main results of this paper, (Theorems \ref{Thm1} and \ref{Thm2} below), provide a step for establishing Conjecture \ref{SSC_3} in the affirmative, and can be summarized in a non-technical way as follows:     

\vspace{.2cm} \noindent {\bf Main Result (non-technical statement).} {\it 
Assume a maximal Cauchy development $\M$ of \eqref{Einstein_eqn} is inextendable as a Lorentzian manifold with metric connection $\Gamma$ locally in $W^{1,p}$, for some $p > 2$. Then, by Theorem \ref{Thm1}, $\M$ is inextendable as a manifold with Lorentzian metric in $C^{0,1}$ and Riemann curvature locally bounded in $L^p$. Moreover, by Theorem \ref{Thm2}, ``generic'' families of extensions with uniform $C^{0,1}$-bounds on metrics and uniform $L^p$-bounds on their curvature are ruled out, assuming no extensions with uniform $W^{1,p}$-bounds on connections exist, $p>4$. 
} \vspace{.2cm}

For proving inextendability at higher regularity is more feasible \cite{LukOh}, the results in this paper provide a key step in validating Conjecture \ref{SSC_3}. Since $W^{1,p}$ is contained in $W^{1,2}$ on bounded domains for $p\geq 2$, Theorems \ref{Thm1} and \ref{Thm2} can readily be combined with results on the Cauchy problem with $L^2$ based Sobolev spaces which, at the forefront of lowest regularities \cite{Klainermann}, fit to our incoming assumptions. The proof of Theorems \ref{Thm1} and \ref{Thm2}, (in Section \ref{Sec_Proof_Thm2}), is based on Blake Temple's and my optimal regularity result for Lorentzian geometry \cite{ReintjesTemple_ell4}, (reviewed in Section \ref{Sec_opt_reg}).

\section{Statement of Results} \label{Sec_results}

Let $\mathcal{O}$ be some open set of generic data. Each datum in $\mathcal{O}$ is a tuple of a Riemannian metric $h_\theta$ and a symmetric $(0,2)$-tensor $K_\theta$ \cite{Choquet},\footnote{Embedding the Cauchy surface $\Sigma$ in spacetime, $K_\theta$ is the second fundamental form on $\Sigma$.} which we represent as $\theta \in \mathcal{O}$  for brevity. To each datum $\theta \in \mathcal{O}$, we consider the maximal Cauchy development $(\mathcal{M}_\theta, g_\theta)$ of the Einstein equations \eqref{Einstein_eqn}, which is a 4-dimensional manifold $\M_\theta$ endowed with a Lorentzian metric $g_\theta$ that solves \eqref{Einstein_eqn}. We assume throughout that the metrics $g_\theta$ are locally in $W^{2,p}$ in each coordinate chart covering $\M_\theta$, for some $p \geq 2$. We denote extensions of $(\mathcal{M}_\theta, g_\theta)$ by $(\mathcal{M}_\theta^\text{ext}, g_\theta)$.             

To state our main theorems, we now introduce what we call {\it bubble extensions}. That is, we assume each extension $(\mathcal{M}_\theta^\text{ext},g_\theta)$ is a manifold $\mathcal{M}_\theta^\text{ext}$ endowed with a Lorentzian metric $g_\theta$, such that $\mathcal{M}_\theta^\text{ext}\setminus \mathring{\mathcal{M}}_\theta$ can be covered by a single coordinate chart $(\Omega_\theta, x_\theta^\mu)$ of $\M_\theta^\text{ext}$. Assume further one can identify all such charts $\Omega_\theta$ with some open set $\Omega \subset \R^4$, (bounded with smooth boundary), on which a coordinate system $x^\mu$ is given, and we assume one can identify all coordinates $x_\theta^\mu$ with $x^\mu$. We refer to each such extension $\mathcal{M}_\theta^\text{ext}$ as a {\it bubble extension} and to $\big(\mathcal{M}_\theta^\text{ext}\big)_{\theta \in \mathcal{O}}$ as a {\it set of bubble extensions}.   We now state the two main results of this paper.

\begin{Thm} \label{Thm1}
Let $(\mathcal{M}_\theta, g_\theta)$ be the maximal Cauchy development of some $\theta \in \mathcal{O}$.  Assume $\mathcal{M}_\theta$ is inextendable in the sense that no bubble extension exists with connection $\Gamma_\theta \in W^{1,p}(\Omega)$, some $p>2$. Then no bubble extension $\mathcal{M}_\theta^\text{ext}$ exists, with H\"older continuous Lorentzian metric $g_\theta\in C^{0,\alpha}(\Omega)$, ($\alpha = 1 - \frac{2}{p}$), metric connection $\Gamma_\theta \in L^{2p}(\Omega)$ and Riemann curvature ${\rm Riem}(\Gamma_\theta) \in L^{p}(\Omega)$.
\end{Thm}

The incoming assumption of Theorem \ref{Thm1} is rather strong, and is indeed incorrect on the interior Kerr spacetime \cite{HawkingEllis}. For this reason it is more relevant to consider families of extensions of generic data $\mathcal{O}$ subject to uniform bounds on their metrics independent of $\theta \in \mathcal{O}$. Our refinement of Theorem \ref{Thm1}, establishing such uniform bounds, requires the stronger assumption that $p>4$, and implies inextendability with Lipschitz continuous metrics with $L^p$ bounded curvature. 

\begin{Thm} \label{Thm2}
Consider a set of generic data $\mathcal{O}$ with maximal Cauchy developments $(\mathcal{M}_\theta, g_\theta)$ for each $\theta \in \mathcal{O}$.  
Assume the Cauchy developments $\mathcal{M}_\theta$ are inextendable in the sense that no set of bubble extensions exists with connections $\Gamma_\theta \in W^{1,p}(\Omega)$, some $p>4$, subject to the uniform bound 
\beq \label{uniform_bound_assumption}
\|\Gamma_\theta\|_{W^{1,p}(\Omega)} \leq C, 
\eeq
for all $\theta \in \mathcal{O}$, for some constant $C>0$ independent of $\theta \in \mathcal{O}$.     
Then no set of bubble extensions $\big(\mathcal{M}_\theta^\text{ext}, g_\theta\big)_{\theta \in \mathcal{O}}$ exists, with Lorentzian metrics $g_\theta\in C^{0,1}(\Omega)$ and connections $\Gamma_\theta \in L^{\infty}(\Omega)$, subject to the uniform bound\footnote{Since metric derivatives and connection components are in one-to-one correspondence by Christoffel's formula, it follows that the $L^\infty$ bound on $g_\theta$ and $\Gamma_\theta$ in \eqref{uniform_bound} is equivalent to a $W^{1,\infty}$ bound on $g_\theta$, which in turn is equivalent to a Lipschitz bound on $g_\theta$, c.f. \cite{Evans}.}   
\beq \label{uniform_bound}
\|g_\theta\|_{L^\infty(\Omega)} + \|\Gamma_\theta\|_{L^{\infty}(\Omega)} + \|{\rm Riem}(\Gamma_\theta)\|_{L^{p}(\Omega)} \leq M,
\eeq
for all $\theta \in \mathcal{O}$, for some constant $M>0$ independent of $\theta$.
\end{Thm}

\subsection{Coordinate based Sobolev norms and spaces}
We denote with $W^{m,p}(\Omega)$ the Sobolev space of functions with weak derivatives up to order $m$ being in $L^p(\Omega)$, ($m\geq 0$, $1\leq p \leq \infty$). We say a tensor or a connection is in $W^{m,p}(\Omega)$ in some coordinate chart $\Omega \subset \R^4$, if all its components are functions in $W^{m,p}(\Omega)$ in their respective coordinate representation. Correspondingly, we take Sobolev norms $\|\cdot \|_{W^{m,p}(\Omega)}$ component-wise on tensors and connections in a coordinate representation, based on a fixed coordinate system $x$, (of bubble extension, unless otherwise stated). For example, the $W^{1,p}$-norm on connection components $\Gamma$ is
\beq  \label{norms}
\| \Gamma \|_{W^{1,p}(\Omega)} \equiv  \|\Gamma \|_{L^p(\Omega)} +  \sum_{\rho=0,...,3} \| \partial_\rho \Gamma \|_{L^p(\Omega)} , 
\hspace{.5cm}
\| \Gamma \|_{L^p(\Omega)} \equiv   \sum_{\sigma,\mu,\nu} \big\|\Gamma^\sigma_{\mu\nu}\|_{L^p(\Omega)}   ,
\eeq
where $\partial_\rho \equiv \frac{\partial}{\partial x^\rho}$ denotes partial differentiation in $x$-coordinates taken component-wise on tenors and connections, and integration is taken with respect to the volume element of the Euclidean metric in $x$-coordinates. Note, $W^{m,p}$ regularity of tensors is preserved under $W^{m+1,p}$ coordinate transformations, and $W^{m,p}$ regularity of connections is preserved under $W^{m+2,p}$ coordinate transformations. However, the value of $W^{m,p}$-norms is coordinate dependent, a problematic issue in Lorentzian geometry circumvented here by restricting consideration to bubble extensions.

\subsection{Strategy of Proof}       
The proof of Theorems \ref{Thm1} and \ref{Thm2} is based on our optimal regularity result in \cite{ReintjesTemple_ell4}. That is, (focusing on Theorem \ref{Thm2}), we assume for contradiction there exist a bubble extension $\big(\mathcal{M}_\theta^\text{ext}\big)_{\theta \in \mathcal{O}}$ with metric connections $\Gamma_\theta \in L^{\infty}(\Omega)$ and curvature ${\rm Riem}(\Gamma_\theta) \in L^p(\Omega)$ subject to the uniform bound \eqref{uniform_bound}. Then the optimal regularity result in \cite{ReintjesTemple_ell4} implies that, locally, there exists a coordinate transformation which maps $\Gamma_\theta$ to optimal regularity, $\Gamma_\theta \in W^{1,p}$ with uniform $W^{1,p}$ bounds, in the new coordinate system. This can then be shown to contradict our assumption in Theorem \ref{Thm2} that no such extension with $\Gamma_\theta \in W^{1,p}$ exists. We give the detailed proofs of Theorems \ref{Thm1} and \ref{Thm2} in Section \ref{Sec_Proof_Thm2}. For this, we state  in Section \ref{Sec_opt_reg} our optimal regularity result in \cite{ReintjesTemple_ell4} (as Theorem \ref{Thm_Smoothing}), and outline its proof based on the {\it Regularity Transformation (RT-) equations}, a system of partial differential equations (PDE's), {\it elliptic} regardless of metric regularity.

\subsection{Remarks}
Our earlier results in \cite{ReintjesTemple_ell2} yield optimal regularity at higher levels of Sobolev regularity, furnishing coordinate transformations which map non-optimal connections in $W^{m,p}$ with Riemann curvature in $W^{m,p}$, ($m\geq1$, $p>4$), to optimal connection regularity $W^{m+1,p}$. By applying this results in \cite{ReintjesTemple_ell2}, one can extend
Theorems \ref{Thm1} and \ref{Thm2} to assert inextendability of maximal Cauchy developments to Lorentzian manifolds with connections of regularity $W^{m,p}$ and Riemann curvature bounded in $W^{m,p}$, under the assumption of inextendability to Lorentzian manifolds with connections of regularity $W^{m+1,p}$, ($m\geq1$, $p>4$).

\section{Optimal regularity in Lorentzian geometry by the RT-equations}  \label{Sec_opt_reg}

We now introduce the optimal regularity result \cite[Thm 2.1]{ReintjesTemple_ell4}, due to Blake Temple and myself, on which the proof of Theorems \ref{Thm1} and \ref{Thm2} is based. To state the theorem, consider a fixed chart $(\Omega,x)$ on a $4$-dimensional manifold $\mathcal{M}$, such that $\Omega_x \equiv x(\Omega) \subset \R^4$, (the image of $\Omega$ under the coordinate map), is open and bounded with smooth boundary.\footnote{Our use here of $\Omega$, as a chart on $\M$, slightly differs from the use of $\Omega \subset \R^4$ in Section \ref{Sec_results}, (where we identified $\Omega \equiv \Omega_x \subset \R^4$), but this ambiguity is irrelevant since our result and methods are local.} Let $\Gamma_x$ denote the collection of components of an affine connection $\Gamma$ in $x$-coordinates, $\Gamma_x\equiv\Gamma^k_{ij}(x)$.  We view $\Gamma_x$ as a matrix valued $1$-form in $x$-coordinates, $(\Gamma_x)^\mu_{\nu} \equiv (\Gamma_x)^\mu_{\nu j} dx^j$. Let $d\Gamma_x$ denote its exterior derivative, $d(\Gamma_x)^\mu_\nu \equiv \partial_i (\Gamma_x)^\mu_{\nu j} dx^i \wedge dx^j$, (using $\mu, \nu$ to denote matrix indices). Writing the Riemann curvature tensor as a matrix valued $2$-form, ${\rm Riem}(\Gamma_x) = d\Gamma_x +\Gamma_x \wedge \Gamma_x$, it follows that the assumption $\Gamma_x \in L^{2p}(\Omega_x)$ and ${\rm Riem}(\Gamma_x) \in L^p(\Omega_x)$ is equivalent to the assumption $\Gamma_x \in L^{2p}(\Omega_x)$ and $d\Gamma_x \in L^p(\Omega_x)$; we henceforth assume the latter. We now state the version of Theorem 2.1 in \cite{ReintjesTemple_ell4} relevant for this paper.

\begin{Thm} \label{Thm_Smoothing}  
Assume $\Gamma_x \in L^{2p}(\Omega_x)$ and $d\Gamma_x \in L^{p}(\Omega_x)$ in $x$-coordinates, for some $p>2$. Let $M>0$ be a constant such that   
\beq \label{bound_incoming_ass}
 \|\Gamma_x \|_{L^{2p}(\Omega_x)} + \|d\Gamma_x \|_{L^p(\Omega_x)} \; \leq \; M.
\eeq    
Then for any point $q\in \Omega$ there exists a neighborhood $\Omega' \subset \Omega$ of $q$, and a coordinate transformation $x \to y$ with Jacobian $J=\frac{\partial y}{\partial x}\, \in W^{1,2p}(\Omega'_x),$ such that the connection components $\Gamma_y$ in $y$-coordinates exhibit optimal regularity $\Gamma_y \in W^{1,p}(\Omega'_y)$, where $\Omega_y \equiv y(\Omega)$. Moreover, $\Gamma_y$ satisfies the uniform bound   
\beq \label{bound_optimal_reg1} 
\|\Gamma_y \|_{W^{1,p}(\Omega'_y)}  \leq C(M) , 
\eeq
and the Jacobian $J$ and its inverse $J^{-1}$, both measured in $x$-coordinates, satisfy
\beq \label{bound_optimal_reg2} 
\|J\|_{W^{1,2p}(\Omega'_x)}  + \|J^{-1}\|_{W^{1,2p}(\Omega'_x)} \leq C(M) , 
\eeq
for some constant $C(M) > 0$ depending only on $\Omega', p$ and $M$. Furthermore,  if $p>4$, and if $\|\Gamma_x\|_{L^{\infty}(\Omega_x)}\leq M$ in addition to the curvature bound \eqref{bound_incoming_ass}, then the Euclidean volume of $\Omega'_x$ is bounded from below by $1/M$.\footnote{The neighborhood $\Omega'$ can be taken as $\Omega' = \Omega \cap B_r(q)$, where $B_r(q)$ is the Euclidean ball of radius $r$ in $x$-coordinates. The radius $r$ is uniform of the order of $1/M$, as long as $\|\Gamma\|_{L^{\infty}(\Omega)}\leq M$.} 
\end{Thm}

Theorem 2.1 in \cite{ReintjesTemple_ell4} applies to general (affine) connections on the tangent bundle of an $n$-dimensional manifold $\M$ with $L^p$ bounded curvature, and extends the classical optimal regularity result of Kazdan-DeTurck \cite{KazdanDeTurck} from Riemannian to Lorentzian metrics and to affine connections. The proof of Theorem 2.1 in \cite{ReintjesTemple_ell4} was a long time coming \cite{ReintjesTemple_geo, ReintjesTemple_ell1, ReintjesTemple_ell2, ReintjesTemple_ell3}, motivated by earlier work on non-optimal Lorentzian metrics of shock wave solutions of the Einstein-Euler equations \cite{Israel,GroahTemple,Reintjes,ReintjesTemple1}, all summarized in the RSPA article \cite{ReintjesTemple_ell6}, (including the extension to vector bundles and Yang-Mills gauge theories in \cite{ReintjesTemple_ell5}).  The main idea for establishing Theorem \ref{Thm_Smoothing} was to derive from the connection transformation law a non-invariant system of {\it elliptic} PDE's on the regularizing Jacobian $J$ as an unknown, an idea motivated by the Riemann-flat condition in \cite{ReintjesTemple_geo}. This idea lead to the formulation of the {\it RT-equations} in \cite{ReintjesTemple_ell1}, whose solutions furnish the regularizing transformation.   

\subsection{Derivation of the RT-equations}
We first give the main steps in the derivation of the RT-equations, and then explain how they furnish optimal regularity. For this, assuming there exists a coordinate transformation with Jacobian $J$ mapping $\Gamma_x$ to $\Gamma_y$ (the connection of optimal regularity), we write the connection transformation law in the form
\beq \label{opt_eqn1}
\Gammati = \Gamma - J^{-1} dJ,
\eeq
where $\Gamma \equiv \Gamma_x$ and $\Gammati^k_{ij} = (J^{-1})^k_\gamma J^\alpha_i J^\beta_j (\Gamma_y)^\gamma_{\alpha\beta}$ is the connection $\Gamma_y$ transformed as a tensor to $x$-coordinates. Then differentiating \eqref{opt_eqn1} by the exterior derivative $d$ and co-derivative $\delta$, \eqref{opt_eqn1} implies after careful organization the following two equations
\begin{eqnarray} 
\Delta \Gammati &=& \delta d\Gamma - \delta\big(dJ^{-1} \wedge dJ\big) + d\delta \Gammati ,     \label{opt_eqn2} \\
\Delta J &=& \delta(J\Gamma) - \langle dJ ; \Gammati \rangle - J \delta\Gammati ,  \label{opt_eqn3}
\end{eqnarray}
where $\Delta \equiv \delta d + d \delta = \partial_{x^0}^2 + ... + \partial_{x^3}^2$ is the Euclidean Laplacian, $\langle \cdot\; ; \cdot \rangle$ is a matrix-valued inner product and $\wedge$ the wedge product on matrix valued differential forms, (see \cite[Ch.3]{ReintjesTemple_ell1} or \cite[Ch.5]{ReintjesTemple_ell4} for detailed definitions).             At the current stage, equations \eqref{opt_eqn2} - \eqref{opt_eqn3} neither appear solvable, nor would it be clear whether a solution $J$ would be a true Jacobian integrable to coordinates, i.e. satisfying ${\rm Curl}(J) =0$. To overcome this obstacle, we view $A\equiv \delta\Gammati$ as a free matrix valued parameter function---a choice which appears plausible by the Riemann-flat condition for optimal regularity in \cite{ReintjesTemple_geo} which only involves $d\Gammati$, but not $\delta\Gammati$. Substituting $A\equiv \delta\Gammati$ in \eqref{opt_eqn2} - \eqref{opt_eqn3}, and viewing $A$ as a new unknown matrix valued function, we next impose on equation \eqref{opt_eqn3} the condition ${\rm Curl}(J) =0$ for integrability, in its equivalent form $d\vec{J} =0$ on the vectorization $\vec{J}^\mu = J^\mu_\nu dx^\nu$ of $J$ so that ${\rm Curl}(J)\equiv d\vec{J}$. After careful organization, involving a fortuitous cancellation by which the regularity of terms in the equations match up, the computations in \cite{ReintjesTemple_ell1} lead to the {\it RT-equations}, the following solvable systems of PDE's which is elliptic regardless of metric signature:
\begin{align} 
\Delta \Gammati &= \delta d\Gamma - \delta \big(d J^{-1} \wedge dJ \big) + d(J^{-1} A ), \label{PDE1} \\
\Delta J &= \delta ( J \Gamma ) - \langle d J ; \tilde{\Gamma}\rangle - A , \label{PDE2} \\
d \vec{A} &= \overrightarrow{\text{div}} \big(dJ \wedge \Gamma\big) + \overrightarrow{\text{div}} \big( J\, d\Gamma\big) - d\big(\overrightarrow{\langle d J ; \tilde{\Gamma}\rangle }\big),   \label{PDE3}\\
\delta \vec{A} &= v.  \label{PDE4}
\end{align}
Equation \eqref{PDE3} on the auxiliary field $A$ results from imposing $d\vec{J}=0$ on \eqref{PDE2}, and one can prove integrablity of $J$ to follow from the coupled equations \eqref{PDE2} and \eqref{PDE3}. The unknowns $(\Gammati,J,A)$ in the RT-equations, together with the given non-optimal connection components $\Gamma$, are viewed as matrix valued differential forms. Arrows denote ``vectorization'', mapping matrix valued $0$-forms to vector valued $0$-forms, (e.g. $\vec{A}^\mu = A^\mu_i dx^i$) and $\overrightarrow{\text{div}}$ is a divergence operation which maps matrix valued $k$-forms to vector valued $k$-forms. The vector $v$ in \eqref{PDE4} is free to impose, representing a ``gauge''-type freedom in the equations, reflecting the fact that smooth transformations preserve optimal connection regularity. The operations on the right hand side are formulated in terms of the Cartan Algebra of matrix valued differential forms based on the Euclidean metric in $x$-coordinates, (see \cite{ReintjesTemple_ell1} for detailed definitions and proofs). 

\subsection{How the RT-equations yield optimal regularity}
We now explain how solutions of the RT-equations furnish the coordinate transformations to optimal regularity, on which the proof of Theorem \ref{Thm_Smoothing} in \cite{ReintjesTemple_ell4} is based. It was because our earlier iteration scheme, for solving the RT-equations at higher regularity in \cite{ReintjesTemple_ell2,ReintjesTemple_ell3}, did not close at the low regularity of $L^p$ connections, due to the non-linear term $d J^{-1} \wedge dJ$ in \eqref{PDE1}, that we eventually discovered internal ``gauge''-type transformations on solutions of the RT-equations \eqref{PDE1} - \eqref{PDE4}, that enabled us to separate off \eqref{PDE1} from the remaining equations. This latter system is what we refer to as the \emph{reduced} RT-equations \cite{ReintjesTemple_ell4}, 
\begin{eqnarray} 
\Delta J &=& \delta ( J \mm \Gamma ) - B , \label{RT_withB_2} \\
d \vec{B} &=& \overrightarrow{\text{div}} \big(dJ \wedge \Gamma\big) + \overrightarrow{\text{div}} \big( J\, d\Gamma\big) ,   \label{RT_withB_3} \\
\delta \vec{B} &=& v'.  \label{RT_withB_4}
\end{eqnarray}
The reduced RT-equations \eqref{RT_withB_2} - \eqref{RT_withB_4} are linear in $(J,B)$, and our iteration scheme (based on Poisson equations) in \cite{ReintjesTemple_ell4} applies to the low regularity of $L^p$ connections with $d\Gamma \in L^p$, and establishes existence of solutions $(J,B)$ in a neighborhood $\Omega'$ of any point, such that $J$ is a point-wise invertable matrix. It is a built-in property of \eqref{RT_withB_2} - \eqref{RT_withB_4} that any solution $J$ is a Jacobian integrable to coordinates, as long that the integrability condition $d\vec{J}\equiv {\rm Curl}(J)=0$ holds on the boundary $\partial\Omega$, (accomplished by our existence theory in \cite{ReintjesTemple_ell4}). That is, combining \eqref{RT_withB_2} with \eqref{RT_withB_3}, a computation shows that $\omega\equiv d\vec{J}$ is a solution of the Laplace equation $\Delta \omega=0$, which together with our boundary data implies that $\omega=0$ throughout $\Omega$. This implies that $J$ is a true Jacobian integrable to coordinates. 

Given now a solution $(J,B)$ of the reduced RT-equation \eqref{RT_withB_2} - \eqref{RT_withB_4} with integrable Jacobian $J$, one recovers a solution $(J,\Gammati,A)$ of the full RT-equations \eqref{PDE1} - \eqref{PDE4} by introducing\footnote{The second and third equation in \eqref{Gammati'} define the ``gauge'' transformations of the RT-equations.}
\beq  \label{Gammati'} 
\Gammati \equiv \Gamma - J^{-1} dJ, 
\hspace{.5cm} 
A \equiv B - \langle d J ; \Gammati \rangle ,  
\hspace{.5cm} \text{and} \hspace{.5cm}
v \equiv v' - \delta \overrightarrow{\langle d J ; \Gammati \rangle},
\eeq
as can be verified by direct computation using \eqref{RT_withB_2} to eliminate uncontrolled terms involving $\delta\Gamma$. From interior elliptic estimates, applied to the first RT-equations \eqref{PDE1}, one can prove that $\Gammati$ is in $W^{1,p}$, a gain of one derivative over $\Gamma$. Defining 
\beq \label{Gamma_y_reverse}
(\Gamma_y)^\gamma_{\alpha\beta} \equiv J_k^\gamma (J^{-1})^i_\alpha  (J^{-1})^j_\beta   \; \Gammati^k_{ij},
\eeq
a comparison of \eqref{Gammati'} with the connection transformation law \eqref{opt_eqn1} implies that $\Gamma_y$ is indeed the transformed connection of optimal regularity, $\Gamma_y \in W^{1,p}(\Omega)$. These are the essential ideas underlying the proof of Theorem \ref{Thm_Smoothing}, which is worked out in full detail at the level of weak solution in \cite{ReintjesTemple_ell4}.

\section{Proof of Theorems \ref{Thm1} and \ref{Thm2}}   \label{Sec_Proof_Thm2}

\subsection{Proof of Theorem \ref{Thm2}}  
We prove Theorem \ref{Thm2} by contradiction, using the optimal regularity result of Theorem \ref{Thm_Smoothing}.  So assume there exist a bubble extension $\big(\mathcal{M}_\theta^\text{ext}\big)_{\theta \in \mathcal{O}}$ with $C^{0,1}$ Lorentzian metrics and metric connections $\Gamma_\theta \in L^{\infty}(\Omega)$ subject to the uniform bound \eqref{uniform_bound}, where $(\Omega,x)$ is the single coordinate chart assumed to cover $\M_\theta^\text{ext}\setminus \mathring{\M}_\theta$ after suitable identification; we denote here with $\Omega_x \equiv x(\Omega) \subset \R^4$ the image of $\Omega$ under the coordinate map following the notation in Section \ref{Sec_opt_reg}. In fact, to apply Theorem \ref{Thm_Smoothing} it suffices to assume the bound 
\beq \label{uniform_bound_proof}
\|\Gamma_\theta\|_{L^{\infty}(\Omega_x)} + \|{\rm Riem}(\Gamma_\theta)\|_{L^{p}(\Omega_x)} \leq M,
\eeq
for some constant $M>0$ independent of $\theta$, some $p>4$. Equation \eqref{uniform_bound_proof} is equivalent to the uniform bound \eqref{bound_incoming_ass} of Theorem \ref{Thm_Smoothing}, as can be shown directly by using H\"older's inequality in combination with the expression for the Riemann curvature tensor as a matrix valued $2$-form, ${\rm Riem}(\Gamma_x) = d\Gamma_x +\Gamma_x \wedge \Gamma_x$.    

So assume \eqref{uniform_bound_proof} and let $q\in \Omega$ be a point on the boundary of $\M_\theta$; (note that this $q\in \Omega$ exists, since $\Omega$ is an open set of $\M_\theta^\text{ext}$ covering $\M_\theta^\text{ext}\setminus \mathring{\M_\theta}$, where $\mathring{\M_\theta}$ is the interior of the maximal Cauchy development $\M_\theta$).   Theorem \ref{Thm_Smoothing} implies that there exists a neighborhood $\Omega' \subset \Omega$ of $q$, (depending on $M$, but independent of $\theta$), such that for each connection $\Gamma_\theta$ there exists a coordinate transformation $x \to y_\theta$ with Jacobian $(J_\theta)^\mu_\nu = \frac{\partial y^\mu_\theta}{\partial x^\nu} \in W^{1,2p}(\Omega'_x)$ such that the connections $\Gamma_{y_\theta}$ in $y_\theta$-coordinates exhibit optimal regularity, $\Gamma_{y_\theta} \in W^{1,p}(\Omega'_{y_\theta})$, where $\Omega'_{y_\theta} \equiv y_\theta(\Omega')$, subject to the uniform bound
\beq \label{proof_eqn1}
\| \Gamma_{y_\theta} \|_{W^{1,p}(\Omega'_{y_\theta})} + \|J_\theta\|_{W^{1,2p}(\Omega'_x)}  + \|J^{-1}_\theta\|_{W^{1,2p}(\Omega'_x)} < C(M),
\eeq
for some constant $C(M)>0$ independent of $\theta$. The norm of the second and third term in \eqref{proof_eqn1} are measured in $x$-coordinates, but the first one is measured in  $y_\theta$-coordinates, the coordinates of optimal regularity for each $\Gamma_\theta$ respectively.    

In order to express \eqref{proof_eqn1} as a uniform bound  suitably measuring the connections $\Gamma_{y_\theta}$ in a single fixed coordinate system, we first write each $\Gamma_{y_\theta}$ in $x$-coordinates by transforming connection components as a scalar function, $\Gamma_{y_\theta}(x) \equiv \Gamma_{y_\theta}(y_\theta(x))$. Now, changing integration from $x$ to $y_\theta$-coordinates, we have
\beq \label{proof_eqn2}
\int_{\Omega'_x} \Gamma_{y_\theta}(x) dx  
= \int_{\Omega'_{y_\theta}} \Gamma_{y_\theta}(y_\theta) \; |{\rm det}(J_\theta^{-1})|\; d y_\theta,
\eeq
which, using that the $L^\infty$-norm is identical in $x$- and $y_\theta$-coordinates under scalar transformations, implies
\begin{eqnarray} \label{proof_eqn3}
\| \Gamma_{y_\theta}(x) \|_{L^p(\Omega'_x)} 
&\leq & \|J_\theta^{-1}\|_{L^\infty(\Omega'_x)} \;  \| \Gamma_{y_\theta}\|_{L^p(\Omega'_{y_\theta})}  \cr
&\leq & C_M \|J_\theta^{-1}\|_{W^{1,2p}(\Omega'_x)} \;  \| \Gamma_{y_\theta}\|_{L^p(\Omega'_{y_\theta})} 
\end{eqnarray}
where we used Morrey's inequality in the last line, and where $C_M>0$ is a constant depending only on $p$ and $\Omega'$, independent of $\theta$.\footnote{Morrey's inequality bounds the H\"older norms of a function $f$ as $\| f\|_{C^{0,\alpha}(\overline{\Omega})}  \leq C_M \|f\|_{W^{1,p}(\Omega)}$, where $\alpha \equiv 1 - \frac{n}{p}$ and $C_M>0$ is a constant depending only on $n$, $p$ and $\Omega \subset \R^n$, c.f. \cite{Evans}.} Similarly, the chain rule implies for differentiation in $x$- versus $y_\theta$-coordinates that
\beq \label{proof_eqn4}
 \partial_{x^\nu} \Gamma_{y_\theta}(x)  =  \sum_{\mu =0,...,3} (J_\theta)^\mu_\nu \; \partial_{y_\theta^\mu}\Gamma_{y_\theta}   ,
\eeq   
which allows us to bound each connection derivative by
\begin{align} \label{proof_eqn5'}
\| \partial_x \Gamma_{y_\theta}(x) \|_{L^p(\Omega'_x)} 
&\leq  C_p\; \| J_\theta\|_{L^\infty(\Omega'_x)}   \;  \| \partial_{y_\theta}\Gamma_{y_\theta} \|_{L^p(\Omega'_x)}  \cr
& \leq  C_M\: C_p\; \| J_\theta\|_{W^{1,2p}(\Omega'_x)}  \; \| \partial_{y_\theta}\Gamma_{y_\theta} \|_{L^p(\Omega'_{x})},
\end{align} 
where $C_p>1$ is a combinatorial constant independent of $\theta$ to account for the summation in \eqref{proof_eqn4}, $\partial_x$ and $\partial_{y_\theta}$ denotes the collection of partial derivatives in respective coordinates (which norms are summed over, c.f. \eqref{norms}), and we applied again Morrey's inequality.  Changing integration to $y_\theta$ coordinates, \eqref{proof_eqn5'} implies 
\begin{align} \label{proof_eqn5}
\| \partial_x \Gamma_{y_\theta}(x) \|_{L^p(\Omega'_x)} 
& \leq  C_M^2 \: C_p\; \| J_\theta\|_{W^{1,2p}(\Omega'_x)} \| J^{-1}_\theta\|_{W^{1,2p}(\Omega'_x)}   \; \| \partial_{y_\theta}\Gamma_{y_\theta} \|_{L^p(\Omega'_{y_\theta})},
\end{align} 
following the reasoning in \eqref{proof_eqn4}. Combining \eqref{proof_eqn3} and \eqref{proof_eqn5}, we can now bound the $W^{1,p}$-norm of $\Gamma_{y_\theta}(x)$ in $x$-coordinates by
\begin{align} \label{proof_eqn6}
\|& \Gamma_{y_\theta}(x)  \|_{W^{1,p}(\Omega'_x)} 
\equiv  \;  \| \Gamma_{y_\theta}(x) \|_{L^p(\Omega'_x)} + \sum_{\nu =0,...,3} \| \partial_{x^\nu}\Gamma_{y_\theta}(x) \|_{L^p(\Omega'_x)}   \cr
& \leq  \ C_M\: C_p\; \| J^{-1}_\theta\|_{W^{1,2p}(\Omega'_x)} \big( 1 + C_M \| J_\theta\|_{W^{1,2p}(\Omega'_x)} \big)  \; \| \Gamma_{y_\theta} \|_{W^{1,p}(\Omega'_{y_\theta})}   .
\end{align}
Using finally \eqref{proof_eqn1} to bound $\| J_\theta^{-1} \|_{W^{1,2p}(\Omega'_x)}$ and $\| \Gamma_{y_\theta} \|_{W^{1,p}(\Omega'_{y_\theta})}$ in \eqref{proof_eqn6}, we obtain the sought-after uniform bound
\begin{eqnarray} \label{proof_eqn7}
\| \Gamma_{y_\theta}(x) \|_{W^{1,p}(\Omega'_x)} 
&\leq &  C_M\: C_p\; \big( 1 + C_M C(M)\big) \;  C(M)^2  \equiv C,
\end{eqnarray}
where both $C>0$ and $\Omega'$ are independent of $\theta$. 

Estimate \eqref{proof_eqn7} contradicts the assumption of Theorem \ref{Thm2}. Namely, the connections $\Gamma_{y_\theta}$ form a set of bubble extensions with $\M_\theta^\text{ext} \equiv \M_\theta \cup \Omega'$, because $\M_\theta^\text{ext} \setminus \mathring{\M}_\theta$ can be covered by $\Omega'$, and since one can identify the coordinates $y_\theta$ with $x$ and $\Omega'_{y_\theta}$ with $\Omega'_x$. Thus, by \eqref{proof_eqn7}, we constructed a set of bubble extensions subject to the uniform bound \eqref{uniform_bound_assumption}, in contradiction to our assumptions in Theorem \ref{Thm2}. We conclude no $C^{0,1}$ Lorentzian metric extension with $L^p$ bounded Riemann curvature, subject to the uniform bound \eqref{uniform_bound}, exists for $p>4$. This completes the proof of Theorem \ref{Thm2}.     
\hfill $\Box$

\subsection{Proof of Theorem \ref{Thm1}}
Let $(\mathcal{M}_\theta, g_\theta)$ be the maximal Cauchy development of some datum $\theta \in \mathcal{O}$, such that no bubble extension of $(\mathcal{M}_\theta, g_\theta)$ exists with connection $\Gamma_\theta$ in $W^{1,p}(\Omega)$, for some $p>2$. Assume for contradiction there exists a bubble extension $\mathcal{M}_\theta^\text{ext}$ with Lorentzian metric $g_\theta\in C^{0,\alpha}(\Omega)$, metric connection $\Gamma_\theta \in L^{2p}(\Omega)$ and Riemann curvature ${\rm Riem}(\Gamma_\theta) \in L^{p}(\Omega)$, where $(\Omega,x)$ is the coordinate chart covering $\M_\theta^\text{ext} \setminus \mathring{\M}_\theta$. Now, by Theorem \ref{Thm_Smoothing}, for any point $q \in \Omega$ on the boundary of $\M_\theta$, there exists a neighborhood $\Omega'$ of $q$, on which a coordinate transformation $x\to y_\theta$ is defined such that $\Gamma_{y_\theta}$ has optimal regularity,  $\Gamma_{y_\theta} \in W^{1,p}(\Omega'_{y_\theta})$. This is a contradiction to our incoming assumption that no such bubble extension exists. This completes the proof of Theorem \ref{Thm1}.                     
\hfill $\Box$

\section*{Conclusion}
The results of this paper, under the assumption that no extensions exist with higher connection regularity $W^{1,p}_\text{loc}$, verify the \emph{Strong Cosmic Censorship conjecture with bounded curvature} in the affirmative. This step towards confirming the conjecture uses elliptic PDE theory, (applied to the RT-equations), to lift the problem from the lowest regularity, (of Lipschitz continuous metrics with bounded curvature), by one level, to regularities more accessible to hyperbolic PDE methods.

\section*{Funding}
M. Reintjes was partially supported by CityU Start-up Grant for New Faculty (7200748) and by CityU Strategic Research Grant (7005839).

\section*{Acknowledgement}
I am thankful to Blake Temple for helpful discussions and encouragement.

\end{document}